\documentclass[pra,aps,showpacs, twocolumn]{revtex4-1}
\usepackage{mathtext}
\usepackage[T2A]{fontenc}
\usepackage[cp1251]{inputenc}
\usepackage{epsf}
\usepackage{psfrag}
\usepackage{graphicx}
\usepackage{amssymb,amsmath}

\begin{document}

\title{Light propagation in a random three-dimensional ensemble of point scatterers in a waveguide: size-dependent switching between diffuse radiation transfer and Anderson localization of light}
\author{A. S. Kuraptsev}
\email[]{aleksej-kurapcev@yandex.ru}
\affiliation{\small Peter the Great St. Petersburg Polytechnic University, 195251, St. Petersburg, Russia}
\author{I. M. Sokolov}
\email[]{ims@is12093.spb.edu}
\affiliation{\small Peter the Great St. Petersburg Polytechnic University, 195251, St. Petersburg, Russia}

\date{\today}

\sloppy



\begin{abstract}
Light transport in a disordered ensemble of resonant atoms placed in a waveguide is found to be very sensitive to the sizes of cross section of a waveguide. Based on self-consistent quantum microscopic model treating atoms as coherent radiating dipoles, we have shown that the nature of radiation transfer changes from Anderson localization regime in a single-mode waveguide to a traditional diffuse transfer in a multi-mode one. Moreover, the transmittance magnitude undergoes complex step-like dependence on the transverse sizes of a waveguide.
\end{abstract}
\maketitle

The transition from extended to localized eigenstates
upon increasing disorder in a quantum or wave system is
named after Philip Anderson who was the first to predict it
for electrons in disordered solids \cite{1}. This transition manifest itself in sharp decrease of the diffusion coefficient for electrons down to zero and associated suppression of the conductivity, thus a conducting medium turns into a dielectric. Actually, the effect of phase transition induced by disorder has a general nature. More recently, it was studied for various types of quantum
particles (cold atoms \cite{2}, Bose-Einstein condensates \cite{3})
as well as for classical waves (light \cite{4,5,6}, ultrasound \cite{7,8}). Anderson localization of light may find
applications in the design of future quantum-information
devices \cite{9}, miniature lasers \cite{10}, and solar cells \cite{11}. Drawing a parallel between Anderson localization predicted for electrons in solids and localization of light in dispersive medium, the analog of metallic phase is the regime of diffuse radiation transfer whereas the analog of non-conducting phase is expressed in Anderson localization of light. The phenomenon of Anderson localization of light can manifest itself in various physical systems, but a special attention is paid to random ensembles of point scatterers (such as cold atomic gases or impurity atoms embedded in a solid transparent dielectric matrix). This is connected with the fact that these objects represent an excellent playground for testing the theory.

In the literature, one can find different signs for Anderson localization of light based on eigenstates diagram (mainly, lifetimes of eigenstates), on the character of radiation trapping, on the inverse participation ratio, on the Ioffe-Regel parameter, on the Thouless number, to name a few. However, these signs represent the necessary conditions for Anderson localization but not sufficient ones. The most reliable criterion comes from the exact definition of localization, namely the transmittance in a stationary mode must exponentially decrease with increasing of the thickness of a sample. There are also some requirements, which have to be met if we want to use this criterion. The first is the absence of any relaxation channels which lead to energy losses of the joint atomic-field system. This circumstance is not a problem in theoretical simulations, but it can represent a critical point for experiments since the existence of such channels permits alternative explanations for exponential behavior of the transmittance \cite{12,13,14}. The second is that a medium has to be optically dense, so the intensity of transmitted light is determined mainly by its incoherent component. Actually, both these requirements are valid not only in the case of transmission criterion but also for all other signs for Anderson localization of light.

By now, it has been understood that Anderson localization of light in a random three-dimensional ensemble of point scatterers without applying external control fields is absent \cite{15}. However, static magnetic field restores Anderson localization of light in a cold-atom gas \cite{16}. The situation essentially changes when we deal with low-dimensional systems. It is known that in two dimensions, there is no true metallic behavior of a disordered electronic system \cite{17}. Thus, reduced dimensionality facilitates the achievement of Anderson localization. This also holds true when understanding Anderson localization in a broad sense, including both metal-insulator transitions and quantum-Hall-type transitions between phases with localized states \cite{18}. Focusing on the Anderson localization of light in atomic ensembles, it is worth to note that the fabrication of low-dimensional ensembles usually assumes their coupling with a cavity or waveguide structures. This imprints the nature of interaction between atoms and electromagnetic field, that, in turn, leads to a modification of interatomic dipole-dipole interaction \cite{19,20,21} and associated cooperative effects \cite{22,23,24,25,26}. Correct description of these effects requires a realistic three-dimensional picture both because of 3D arrangement of atoms and, which is more fundamental, because of vector nature of electromagnetic field. Moreover, we have recently shown
that polarization effects play a crucial role in spontaneous decay of an excited atom in a single-mode waveguide \cite{27}.

In this Letter we report the results of fully three-dimensional calculation of light intensity transmitted through a random ensemble of point scatterers in a waveguide. We show that the nature of light transport dramatically depends on the transverse sizes of a waveguide. Thus, when a waveguide is single-mode with respect to resonant transition wavelength, the regime of Anderson localization is realized even for an arbitrarily low  atomic density. An increase in the transverse size turns a single-mode waveguide to a multimode one, that, in turn, instantly cancels Anderson localization and restores classical diffuse radiation transfer.

We consider an ensemble of $N\gg1$ identical two-level
atoms at random position $\{\textbf{r}_i\}$ inside a waveguide, see Fig. 1. The resonant frequency $\omega_0$ of atoms defines the
natural length scale $1/k_0=c/\omega_0$, where $c$ is the vacuum
speed of light. The ground state $|g_i\rangle$ of an isolated atom $i$ is
nondegenerate with the total angular momentum $J_{g}=0$,
whereas the excited states $|e_i\rangle$ is threefold degenerate with
$J_{e}=1$ and natural free
space linewidth $\gamma_0$. The three degenerate substates $|e_{i,m_J}\rangle$ correspond to
the three possible projections $m_{J}=0,\pm 1$ of the total
angular momentum $\textbf{J}_{e}$ on the quantization axis $z$. For convenience, let us choose the $z$ axis
coinciding with the axis of a waveguide.

\begin{figure}\center
	\includegraphics[width=7cm]{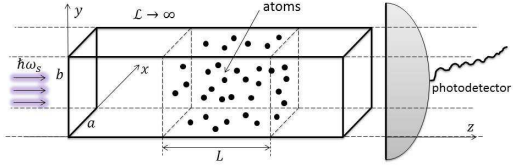}
	\caption{\label{fig:one}
			Sketch of the waveguide and the atomic ensemble inside it.}\label{f1}
\end{figure}

The cross section of the waveguide is assumed to be rectangular having the sizes $a$ and $b$. Atomic ensemble occupies a whole cross section of a waveguide and has the longitudinal length $L$ along $z$ axis. Both input and output of a waveguide are far-remote from the edges of atomic medium (corresponding separations are much larger than $L$). Atomic ensemble is illuminated by stationary probe radiation, which is considered to be monochromatic with the frequency $\omega_s$. Transmitted radiation is measured by a photodetector, which absorbs the whole output signal integrated over the area of cross section indifferent to polarization state.

To describe the stationary regime of atomic excitation induced by external radiation within the framework of consistent quantum-mechanical treatment, we use the following technique. Let us suppose that probe radiation is created as a result of spontaneous emission of some remote atom in a waveguide, which has the same level structure as atoms from an ensemble but different resonance transition frequency $\omega_s$ and narrow linewidth $\gamma_{s}$ (hereafter we call it as "source-atom"). It allows us to consider the following initial conditions: only source-atom is excited whereas all other atoms forming an ensemble are in their ground state; electromagnetic field is in the vacuum state at $t=0$. Thus, the problem considered here can be formally reduced to the problem of collective spontaneous decay in a waveguide, which we studied previously \cite{27}.

Assuming the walls
of a waveguide to be perfectly conductive (i.e., neglecting the
absorption), the dynamics of atomic-field system can be treated on the basis of the non-steady-state Schrodinger
equation with the following Hamiltonian \cite{28}:
\begin{eqnarray}\label{lett1}
  \widehat{H}&=&\sum_{i=1}^{N+1}\sum_{m_{J}=-1}^{1}\hbar\omega_{0}|e_{i,m_{J}}\rangle\langle e_{i,m_{J}}| \nonumber \\
  &+&\sum_{\textbf{k},\alpha}\hbar\omega_{k}\left(\widehat{a}_{\textbf{k},\alpha}^{\dagger}\widehat{a}_{\textbf{k},\alpha}+\frac{1}{2}\right)-\sum_{i=1}^{N+1}\widehat{\textbf{D}}_{i}
  \cdot\widehat{\textbf{E}}\left(\textbf{r}_{i}\right) \nonumber \\
  &+&\frac{1}{2\epsilon_{0}}\sum_{i\neq j}^{N+1}\widehat{\textbf{D}}_{i}\cdot\widehat{\textbf{D}}_{j}\delta\left(\textbf{r}_{i}-\textbf{r}_{j}\right),
\end{eqnarray}

where the first two terms correspond to noninteracting
atoms and the electromagnetic field in an empty waveguide, respectively, the
third term describes the interaction between the atoms and
the field in the dipole approximation, and the last, contact
term ensures the correct description of the electromagnetic
field radiated by the atoms \cite{28}. In Eq. (\ref{lett1}), $\widehat{a}_{\textbf{k},\alpha}^{\dagger}$ and $\widehat{a}_{\textbf{k},\alpha}$ are the
operators of creation and annihilation of a photon in the corresponding mode, $\omega_{k}$ is the photon frequency, $\widehat{\textbf{d}}_{i}$ is the dipole operator of the atom $i$, $\widehat{\textbf{E}}\left(\textbf{r}\right)$ is the electric displacement vector in a waveguide, and $\textbf{r}_{i}$ is the position of the atom $i$. The vacuum reservoir is also included in the atomic-field system described by the Hamiltonian represented in Eq. (\ref{lett1}).

Formally solving the Schrodinger equation for the joint system, which consists of $N+1$ atom ($N$ atoms of an ensemble + source-atom) and the electromagnetic field, and restricting
ourselves by the states containing no more than one photon (i.e. neglecting nonlinear effects), one obtains a system of equations for the amplitudes $b_e$ of one-fold atomic excited states with the coupling
between atoms caused by dipole-dipole interaction. For Fourier components $b_{e}(\omega)$ we have
(at greater length see \cite{29})
\begin{equation}\label{2}
\sum_{e'}\bigl[(\omega-\omega_{e})\delta_{ee'}-\Sigma_{ee'}(\omega)\bigl]b_{e'}(\omega)=i\delta_{e s}.
\end{equation}
The index $s$ as well as the indexes $e$ and $e'$ contain information both about the number of atom and about specific atomic sublevel excited in the corresponding state.

The matrix $\Sigma_{ee'}(\omega)$ describes both spontaneous decay and photon exchange between the atoms. It is connected with the Green's matrix $G_{ee'}(\omega)$ by a simple relation $\Sigma_{ee'}(\omega)=(-\gamma_{0}/2)G_{ee'}(\omega)$. The explicit expressions for the elements of the Green's matrix corresponding to a waveguide were derived in \cite{27}.

By the inverse Fourier transform, we get the dynamics of the quantum amplitudes in a time domain, $b_{e}(t)$. Further, we go to the stationary limit of atomic excitation by external unaffected light source. For this, we should remove the reverse influence of the atomic ensemble on the source-atom and consider $\gamma_{s}\rightarrow 0$ and $t\rightarrow \infty$ assuming $\gamma_{s}t\ll 1$. This leads to the following expression:
\begin{equation}\label{3}
b_{e}(t)=\exp(-i\omega_{s}t)\sum_{e'\neq s}R_{ee'}(\omega_{s})\Sigma_{e's}(\omega_{s}),
\end{equation}
where $R_{ee'}(\omega)$ is a resolvent operator of the considered multi-atomic ensemble, which is defined as
$R_{ee'}(\omega)=[(\omega-\omega_{e})\delta_{ee'}-\Sigma_{ee'}(\omega)]^{-1}$.

The transmission coefficient is defined as follows: $T=I/I_{0}$, where $I$ is the intensity of transmitted light and $I_{0}$ is the intensity of incident light; both intensities are averaged over the area of cross section of a waveguide. Transmitted light intensity can be calculated in a straightforward manner, as it was done in Refs. \cite{30,31}, or it can be simulated on the basis of the alternative method, which considers the so-called "atom-detector". The idea of this alternative approach is taking into consideration an imaginary elusive "atom", which sense radiation emitted by the environment medium but do not re-emits photons. So, it works as a point detector. It is also important here that atom-detector must perceive any kind of polarization of electromagnetic waves with equal susceptibility. The intensity of light at the point of atom-detector is proportional to its excited state population. Thus, the calculation of the transmission coefficient, $T$, is reduced to the calculation of the population of excited state of atom-detector: consider the presence of atomic ensemble to find $I$ and its absence to get $I_{0}$.

Figure 2 shows the transmission coefficient depending on the length of atomic sample, $T(L)$, for both cases of single-mode waveguide, Fig. 2(a), and multimode one, Fig. 2(b). The transverse sizes
of a waveguide were chosen as $a=4$, $b=2$ when plotting Fig. 2(a) and $a=b=8$ when plotting Fig. 2(b). The mode composition of a waveguide can be easily justified on the basis of well-known expression for cutoff frequencies of different modes, $\omega_{c}=c\sqrt{(m\pi/a)^2+(n\pi/b)^2}$, where $m$ and $n$ are transverse indexes of given mode. Thus, in the case of $a=4$, $b=2$, only $\text{TE}_{10}$ mode can propagate in a waveguide at long distances as oscillating wave at the transition frequency, while in the case of $a=b=8$, there are 10 such modes: $\text{TE}_{10}$, $\text{TE}_{01}$, $\text{TE}_{20}$, $\text{TE}_{02}$, $\text{TE}_{11}$, $\text{TE}_{12}$, $\text{TE}_{21}$, $\text{TM}_{11}$, $\text{TM}_{12}$, $\text{TM}_{21}$.

\begin{figure}\center
	\includegraphics[width=7cm]{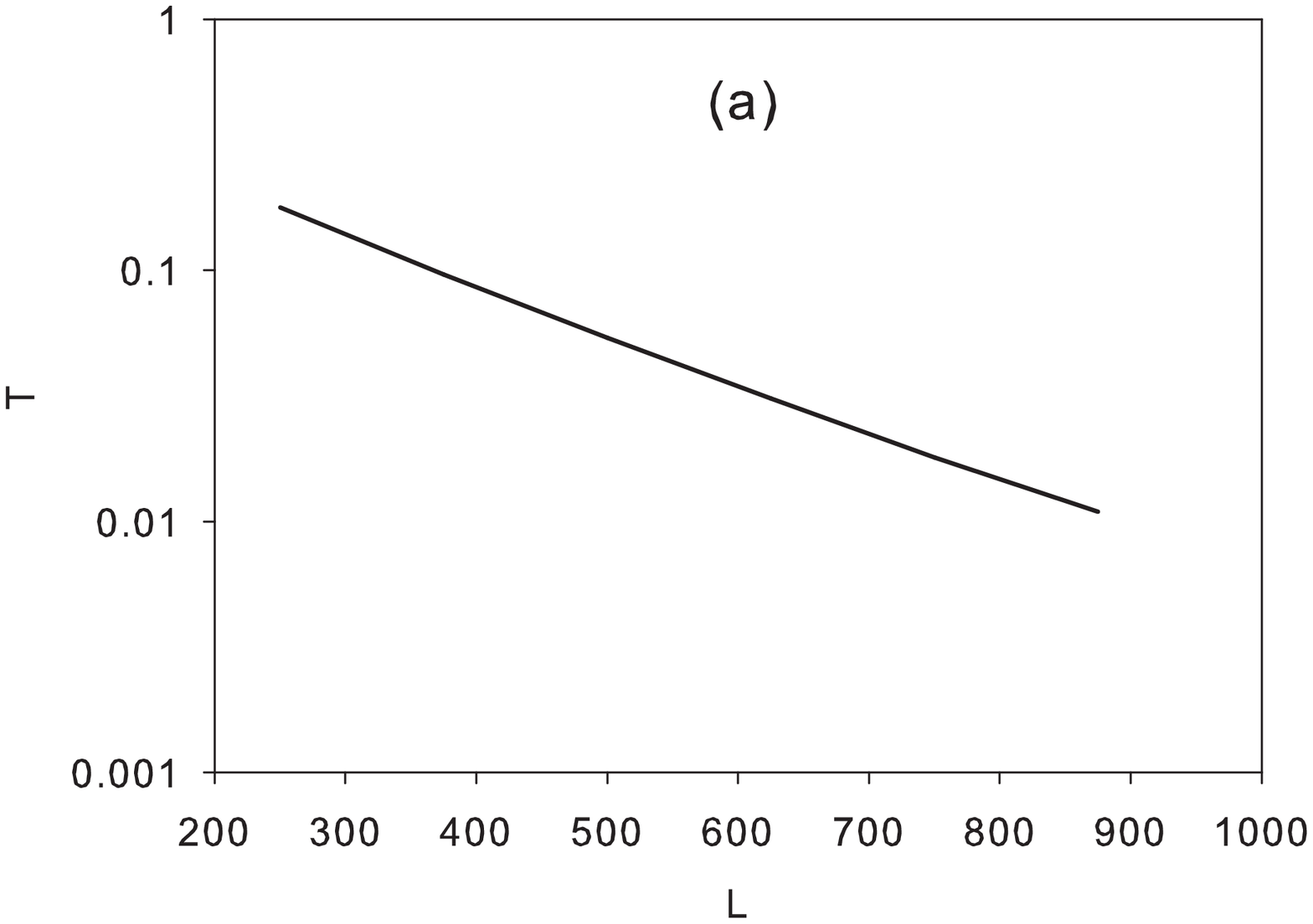}
	\includegraphics[width=7cm]{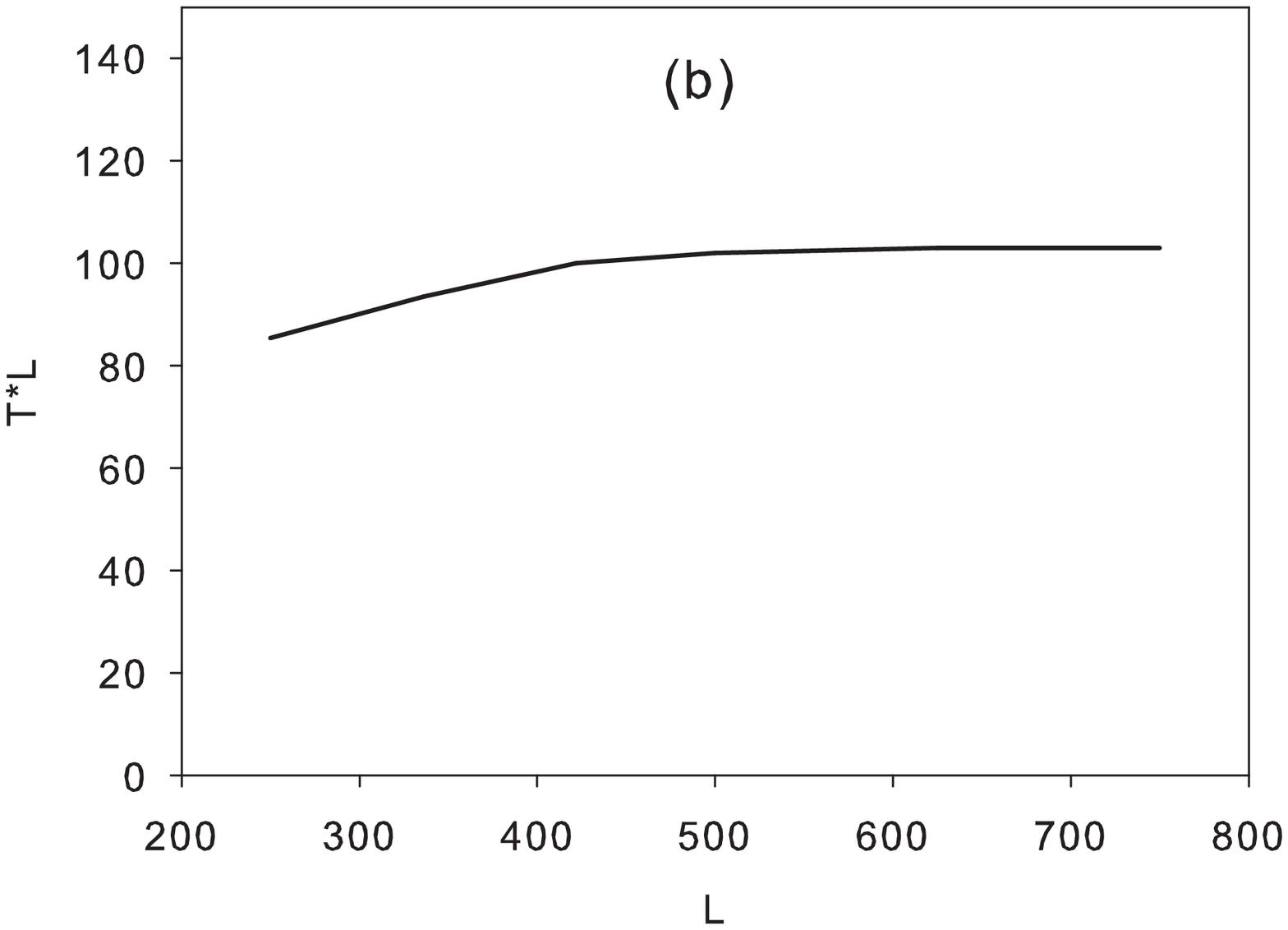}\\
	\caption{\label{fig:two}
			Transmission depending on the length of atomic medium in a waveguide. $n=2\times 10^{-3}$, $\Delta=1$. (a) $a=4$, $b=2$; (b) $a=b=8$.}\label{f2}
\end{figure}

Atomic density is chosen sufficiently small, $n=2\times 10^{-3}$, so that cooperative effects in free space are negligible and, a fortiori, Anderson localization of light in free space ensemble is impossible. The detuning of the probe frequency related to the resonant frequency of atomic transition, $\Delta=\omega_{s}-\omega_{0}$, is chosen $\Delta=1$. Under given condition, the mean free path of a photon can be estimated on the basis of its free space value,
\begin{equation}\label{4}
l_{ph}=\frac{1}{n\sigma_0}\times\frac{\Delta^{2}+(\gamma_{0}/2)^{2}}{(\gamma_{0}/2)^{2}},
\end{equation}
where $\sigma_0$ is single-atom resonant cross section, $\sigma_0=6\pi k_{0}^{-2}$. Substituting the given values of $n$ and $\Delta$ in Eq. (\ref{4}), we get $l_{ph}\approx 133$. So, in order to study incoherent light transfer, we should consider the values of $L$ several times larger than this estimation. That is exactly we do when plotting Fig. 2.

Fig. 2(a) shows that in a single-mode waveguide, the dependence $T(L)$ is exponential, $T(L)=T_{0}\exp(-L/l_{loc})$. This indicates on the regime of Anderson localization of light, the parameter $l_{loc}$ means the localization length and $T_{0}$ is normalization parameter. For given parameters, $T_{0}=0.51$ and $l_{loc}=224$. With increasing in transverse sizes of a waveguide, $a$ and $b$, both parameters $T_{0}$ and $l_{loc}$ undergo complex step-like changes. Sharp steps correspond to the changes in the mode compositions of a waveguide, e.g. when single-mode waveguide turns out to two-mode. With further increasing in the transverse sizes, the nature of transmittance scaling itself changes and ceases to be purely exponential. Thus, in the case of multi-mode waveguide, the dependence $T(L)$ becomes close to hyperbolic, $T(L)\propto1/L$, see Fig. 2(b). This indicates on the regime of classical diffuse radiation transfer.

In conclusion, we have calculated the transmission of disordered atomic ensemble in a waveguide on the basis of self-consistent quantum microscopic treatment taking into account tree-dimensional arrangement of atoms and vectorial nature of electromagnetic field. We have found that the nature of light transport essentially depends on the transverse sizes of a waveguide. Single-mode waveguide with small transverse sizes exhibits Anderson localization of light, which manifests itself in exponential decrease of the transmittance with increasing in the thickness of atomic sample. An increase of the transverse sizes breaks exponential law, so that in a few-mode waveguide we observe complicated dependence of the transmission on the thickness due to complex interplay between Anderson localization of light and diffuse radiation transfer. With further increasing of the transverse sizes of a waveguide, the transmission scaling obeys hyperbolic law, which indicates on the regime of diffuse transfer in a multi-mode waveguide.

\section*{Acknowledgments}
A.S.K. appreciates financial support from the Foundation for the Advancement of Theoretical Physics and Mathematics ''BASIS''.
The results of the work were obtained using computational
resources of Peter the Great Saint-Petersburg Polytechnic
University Supercomputer Center (http://www.scc.spbstu.ru).

\end{document}